\documentclass[12pt,a4paper,normalheadings,headsepline,headinclude,bibtotoc,fleqn,review]{article}
\usepackage[latin1]{inputenc}
\usepackage{amsmath}
\usepackage{booktabs}
\usepackage{array}
\usepackage{amsthm}
\usepackage{dcolumn}
\usepackage{endnotes}
\usepackage{units}
\usepackage{marvosym}
\usepackage[lines=10]{geometry}
\usepackage{longtable}
\usepackage{rotating}
\usepackage{expdlist}
\usepackage{geometry}
\usepackage{floatrow}
\usepackage{pgfplots}
\usepackage{tikz}
\usetikzlibrary{shapes,snakes}
\pgfmathdeclarefunction{gauss}{3}{
	\pgfmathparse{1/(#3*sqrt(2*pi))*exp(-((#1-#2)^2)/(2*#3^2))}%
}
\usetikzlibrary{decorations.pathreplacing,calligraphy,backgrounds}
\pgfplotsset{compat=1.16}

\pgfmathsetmacro\valueA{gauss(2.25, 2.25, 0.7)} 

\usepackage{wrapfig}
\usepackage{expdlist}
\usepackage{amssymb}
\usepackage{gloss}
\usepackage{nomencl}
\usepackage{natbib}

    \makegloss

\renewcommand{\thesection}{\arabic{section}}

\usepackage{fancyhdr} 
\begin{document}

\begin{center}\huge{Games with Planned Actions and Scouting}\footnote{I thank Dominik Grafenhofer, and Manuel Mago, for comments on an earlier draft.}
\end{center}

\begin{center} \textit{Wolfgang Kuhle}\\ \textit{Corvinus University of Budapest, Hungary, E-mail wkuhle@gmx.de\\ MEA, Max Planck Institute for Social Law and Social Policy, Munich, Germany}\end{center} 

\noindent\emph{\textbf{Abstract:} We study games in which every action requires planning and preparation. Moreover, before players act, they can revise their plans based on partially revealing information that they receive on their adversary's preparations. In turn, we examine how players' information over each others' planned actions influences winning odds in matching pennies games, and how it incentivises the use of decoys, deception, and camouflage. Across scenarios, we emphasize that the decomposition of an action into (i) a preparation to act and (ii) the execution of the action, allows to analyze one-shot simultaneous-move games, where players partially observe each others' contemporaneous actions.}

\vspace{.5cm}

\textit{``I'm not kidding, he would go into his rocking motion, his same routine and just as he was about to toss the ball, he would stick his tongue out. And it would either be right in the middle of his lip or it'd be to the left corner of his lip. If he's serving in the deuce court and he put his tongue in the middle of his lip, he was either serving up the middle or to the body. But if he put it to the side, he was going to serve out wide."} Andre Agassi on breaking Boris Becker's serve.\\

\textit{``I used to go home all the time and just tell my wife, it's like he reads my mind. Little did I know, [he was] just reading my tongue."} Boris Becker.

\section{Introduction}\label{Introduction}

In games, such as tennis, soccer, football, or in armed conflict, players carefully position themselves before they act. Moreover, to inform these preparations, sports team and modern armies devote substantial resources, such as drones, satellites, and lip-readers,\footnote{See, e.g., \citet{Lip2001}, for a discussion on how ``Lip reading is a tactic some coaches and scouts are increasingly employing to capture another team's signals, and in turn, anticipate what play is coming." Likewise, consider the Republican's efforts to obtain the Democrats' campaign plans from the Watergate complex.} to scout and decipher their adversary's plans and preparations.



To incorporate the preparations which players undertake before they act, play in our baseline model involves three stages. In the first stage, both players choose which action to prepare. In the second stage, both players receive noisy information over their adversary's preparations. Finally, given the information about their adversary's planned action, and given their own planned action, players choose whether to execute or to revise their plans. Once neither player chooses to make further revisions, plans are executed and payoffs materialize.

\textit{Related literature:} We model the scouting and planning of actions in the context of matching pennies games, which are widely used to study sports, such as tennis, soccer, baseball, or armed conflict.\footnote{\citet{walker2001minimax} use the matching pennies mixed equilibrium to study tennis, where players can serve ``left" or ``right." Likewise, army commanders face similar rock-paper-scissors situations, where fielding tank units yields a victory against an enemy that fields air defense units, while air defense units defeat planes, and planes defeat tanks... . Related, see \citet{schelling1980strategy} for the use of mixed strategies in conflict.} In turn, we show that players' scouting and positioning efforts change equilibrium play and, in particular, winning odds. 

The present paper is related to the literature on (i) learning in games and (ii) conjectural equilibrium.\footnote{See \citet{Bat97}, \citet{Rub94}, or \citet{Ang06}. Likewise, the literature on learning in dynamic games, e.g. \citet{Cha99}, \citet{Pav07}, \citet{Fra12}, emphasizes that players can only learn from each others' past rather than future or contemporaneous actions.} Among others, \citet{Ang06}, p. 1730, argue that ``it is naive" to assume that players can observe each others' actions in simultaneous move games. Likewise, \citet{Rub94} and \citet{grafenhofer2022observing} encounter the similar difficulties, and focus on ``steady states," in which players observe, and repeat, other players' past actions. In the present paper we argue that games, where players' monitor each others' preparations to act, avoid the ``chicken egg" type of problems, between signal and action, that emerge in standard one-shot simultaneous move games where players observe each other.

 Finally, \citet{Rub89}, \citet{Car93}, \citet{Izm10}, \citet{Ste11}, \citet{Kuh15}, \citet{Gra16}, \citet{Ber16}, \citet{Bin01} and \citet{grafenhofer2022observing} have recently emphasized the role of asymmetric information over fundamentals, the role of heterogeneous priors, strategic information revelation, learning, as well as environments where agents receive signals over each others' information. Taking this view, the present paper adds information over planned actions to a literature that studies how different sources of information impacts equilibrium play. 
 
Section \ref{Model} describes the baseline model, where actions are only executed once both players decide that they do not want to revise their plans any further. That is, in the interpretation of an armed conflict, we study games where both commanders, having positioned their troops, accept the engagement.\footnote{In the literature on war practices during the antique epoch, this is known as a ``pitched battle."}
Section \ref{Initiative} extends the model, and assumes that one player has ``the initiative." That is, one player can force the execution of plans, e.g. by serving the ball, in an effort to catch the other player ``on the wrong foot." Section \ref{Decoy} is devoted to the use of decoys, such as inflatable rubber tanks, which players use to offset each others' scouting efforts. Section \ref{Discussion} concludes.

\section{Model}\label{Model}

The game's payoffs in Table 1, are common knowledge among players.
\begin{table}[h]\label{table:kysymys}
			\begin{center}
				\begin{tabular}{cc}
					& Player 2 \\
					Player 1 &
					\begin{tabular}{c|cc}
						& A & B  \\
						\hline
						A & 1;-1 & -1;1\\
						B & -1;1  & 1;-1 \\
					\end{tabular}
					\\ \label{tab:1}
\caption{Payoffs.}
				\end{tabular}
			\end{center}
\end{table}\\
Each \textbf{iteration} of the game involves three stages:
\begin{enumerate}
    \item \textit{Planning/Preparation Stage:} Players $i=1,2$ choose probabilities $\alpha_{i}$ and $1-\alpha_i$ with which they prepare actions $A_i$ or $B_i$.
    
    \item \textit{Intelligence and Scouting Stage:} Once players have chosen their planned actions, each player receives a signal regarding the other player's planned action. We denote the signals that Player 1 receives over Player 2's planned action by $\hat{A}_1$ and $\hat{B}_1$, respectively. Moreover, we assume that Player 1's signal over Player 2's planned action is correct with probability $p>1/2$, such that $P(\hat{A}_1|A_2)=P(\hat{B}_1|B_2)=p$ and $P(\hat{B}_1|A_2)=P(\hat{A}_1|B_2)=1-p$. Likewise, Player 2 receives signals $\hat{A}_2$ and $\hat{B}_2$ over Player 1's planned actions, which are correct with probability $q>1/2$, such that $P(\hat{A}_2|A_1)=P(\hat{B}_2|B_1)=q$ and $P(\hat{B}_2|A_1)=P(\hat{A}_2|B_1)=1-q$.
    
    \item \textit{Execution/Revision of Plans:} Given their planned action and given their signal over the other player's planned action, each player chooses whether to revise or to execute his plan. If at-least one player chooses to revise his plans, the game is reset to the \textit{Planning Stage}.\footnote{That is, we study a pitched battle, where commanders, after they positioned their troops, accept each others' offer to fight the battle. We discuss the alternative case, where one player can force the execution of plans, even though the other player would wish to revise his plans, in Section \ref{Initiative}.} That is, \textbf{a new iteration of the game starts, and the game is reset to the Planning/Preparation Stage}. On the contrary, if neither player wants to make further revisions, planned actions are executed, and payoffs materialize.
\end{enumerate}
Finally, we assume that players choose to execute their plans whenever they cannot expect that a revision of their plans, will bring a plan-signal configuration that offers them higher winning odds in the future.\footnote{This assumption ensures that players do not keep on revising their plans indefinitely. This assumption is implicit in the basic matching pennies game, where players have to play despite the fact that expected payoffs are zero, while the payoff's variance is positive. Alternatively, this assumption corresponds to time limits, which force players to move in games such as tennis, soccer, or baseball.}

\subsection{Equilibrium} 
To characterize equilibrium play, we focus on the probabilities with which signal-plan combinations are drawn that induce both players to execute, rather than to revise, their plans: 
\begin{eqnarray}
   && P(A_1,A_2)=P(\hat{A}_1,\hat{B}_2|A_1,A_2)\alpha_1\alpha_2= p(1-q) \alpha_1\alpha_2 \label{equilibrium1} \\
   && P(B_1,B_2)=P(\hat{B}_1,\hat{A}_2|B_1,B_2)(1-\alpha_1)(1-\alpha_2)= p(1-q)(1-\alpha_1)(1-\alpha_2) \label{equilibrium2} \\
   && P(A_1,B_2)=P(\hat{A}_1,\hat{A}_2|A_1,B_2)\alpha_1 (1-\alpha_2)= (1-p)q\alpha_1 (1-\alpha_2)
    \label{equilibrium3}\\
   && P(B_1,A_2)=P(\hat{B}_1,\hat{B}_2|B_1,A_2)(1-\alpha_1)\alpha_2= (1-p)q (1-\alpha_1)\alpha_2     \label{equilibrium4}
\end{eqnarray}
Equation (\ref{equilibrium1}) reflects that signals $\{\hat{A}_1,\hat{B}_2\}$ incentivize players 1 and 2 to execute plans $\{A_1,A_2\}$. That is, Player 1 is willing to execute planned action $A_1$, given that he receives a signal $\hat{A}_1$, which indicates that Player 2 plans to play $A_2$. Likewise, Player 2 is willing to execute plan $A_2$ conditional on receiving signal $\hat{B}_2$. That is, both players choose to execute plans once they receive signals that indicate a winning position. 

The remaining equations (\ref{equilibrium2})-(\ref{equilibrium4}) indicate that signals $\{\hat{B}_1,\hat{A}_2\}$ incentivize players to execute plans $\{B_1,B_2\}$. Signals $\{\hat{A}_1,\hat{A}_2\}$ incentivize players to execute plans $\{A_1,B_2\}$. Signals $\{\hat{B}_1,\hat{B}_2\}$ incentivize players to execute plans $\{B_1,A_2\}$. Finally, we note that all other signal-plan combinations will not be played in the baseline model as at least one player has an incentive to revise his plans.\footnote{That is, for example, signals $\{\hat{B}_1,\hat{A}_2\}$ in combination with (planned) actions $\{A_1,B_2\}$, induce Player $1$ to revise, rather than to execute, his plans}

 Using equations (\ref{equilibrium1})-(\ref{equilibrium4}) in Appendix \ref{Appendix A}, we find that the probabilities $\alpha_i,1-\alpha_i$, with which players $i=1,2$ plan actions $A_i$ and $B_i$, are: 
 \begin{eqnarray}
   \alpha_1=\alpha_2=1/2 \label{alpha}
\end{eqnarray}
 Finally, summing over the probabilities in equations (\ref{equilibrium1})-(\ref{equilibrium4}), yields the probability $\gamma$, with which players execute, rather than revise, their plans at any given iteration of the game:
\begin{eqnarray}
&&\gamma=p(1-q) \alpha_1\alpha_2+p(1-q) (1-\alpha_1)(1-\alpha_2)+(1-p)q \alpha_1 (1-\alpha_2)+(1-p)q (1-\alpha_1)\alpha_2\nonumber \\
        &&\quad=\frac{1}{2}(p(1-q)+(1-p)q)\label{gamma}
\end{eqnarray}
Equations (\ref{equilibrium1})-(\ref{gamma}) characterize equilibrium play in a matching pennies game where players' partially observe each others' plans and actions. To interpret the present game, we now study how play differs from standard matching pennies mixed strategies equilibrium:

\paragraph{Motivation to Act:} The decision to execute a particular action is always motivated by information that players receive on the adversary's likely course of action. That is, players screen for favorable engagements: every time they act, they do so based on the intelligence that they gathered. On the contrary, in the standard matching pennies game's mixed equilibrium, players choose actions in line with a $50-50$ coin toss. 

Winning odds in basic matching pennies are $50-50$. In the present game, equations (\ref{equilibrium1})-(\ref{alpha}) indicate that, for any values $p$ and $q$, players plan and execute actions $A$ and $B$ with equal probability. However, the same is not true for the winning odds, which depend on players' ability, $p$ and $q$, to monitor their adversaries' preparations. In turn, the probabilities $\pi_1,\pi_2$ with which players 1 and 2 win the game are:
\begin{eqnarray}
    &&\pi_1=\frac{P(A_1,A_2)+P(B_1,B_2)}{\gamma}=\frac{p(1-q)}{(1-p)q+p(1-q)}\label{odds1}\\
    &&\pi_2=1-\pi_1 \label{odds2}
\end{eqnarray}
Moreover, we can define a win ratio as:
\begin{eqnarray}
   && \rho_1=\frac{p(1-q)}{q(1-p)} \label{ratio}
\end{eqnarray}
Using (\ref{odds1})-(\ref{ratio}), setting p=3/4 and q=2/3, the chances for players 1 and 2 to win are 3/5 and 2/5 respectively. That is, a small advantage in the precision with which Player 1 anticipates Player 2's actions correctly gives that player a considerable advantage.\footnote{Put differently, the win ratio $\rho_1$ for Player 1 is 9/6. At the same time, the ratio of the precisions with which players correctly anticipate each others' actions is only 9/8. For $p=q$, the win ratio was equal to one, and thus also equal to the ration between the precisions $p/q$.}

Taking derivatives, equations (\ref{odds1})-(\ref{ratio}) also indicate that players' expected returns from playing the game are monotonously increasing in their signal quality. Moreover, one player's expected marginal return to an increase in signal quality, is low (high) if the other player's signal quality is high (low).

\paragraph{Myopic/Naive Players:} Agents, who do not understand that their adversary uses his own information to screen for favorable engagements, will expect to win with a probability of $p>1/2$ and $q>1/2$, respectively. That is, myopic players will overestimate the probability to win an engagement. Put yet differently, players who do not fully understand the game's equilibrium, and thus the winning odds (\ref{odds1})-(\ref{odds2}), see greater promise in fighting a battle than those who compute winning odds rationally.

\paragraph{Likelihood of Conflict:} In each iteration of the game, the probability that plans and signals are such that players are willing to act is $\gamma<1$. In turn, the probability that players revise plans, starting a new iteration of the game, is $1-\gamma$. Hence, the likelihood that $T$ iterations elapse without an engagement, is $(1-\gamma)^T$.

The probability of an engagement, $\gamma$, is maximized when $p=q=1/2$, i.e., when the information over each others' plans is weakest. On the contrary, in the case where information is perfectly revealing, $p=q=1$, we have $\gamma=0$, and players will never find a position where they are both willing to execute their planned actions. That is, once we interpret our game as one of conflict, the present result indicates that good intelligence reduces the likelihood of conflict at each iteration of the game.

\paragraph{Conjectural Equilibrium in Simultaneous Move Games:} The present model assumes that (i) every action requires preparation and (ii) that these preparations are being monitored such that (iii) players partially know each others' contemporaneous actions once they decide to execute their plans. The current approach thus avoids the ``chicken egg" type of problems, between signal and action, that emerge in standard simultaneous move games where players observe each other.\footnote{See \citet{Ang06}, p. 1730, who argue that ``it is naive" to assume that players can observe each others' actions in simultaneous move games, and \citet{Rub89} for a discussion of the difficulties that prevent the study of one-shot simultaneous move games where players observe each others' contemporaneous actions.} The current modeling approach thus does not require the steady state assumption that \cite{grafenhofer2022observing} and \citet{Rub94} make, when they study conjectural equilibria in simultaneous move games. 

\section{Extensions:} Our baseline model suggests a number of extensions. First, Section \ref{Initiative} studies games where one player can force the execution of plans in an effort to catch the other player ``off guard." Second, Section \ref{Decoy} examines games (i) where some plans are harder to read than others and (ii) explores how players can use decoys to degrade their adversaries information. 

\subsection{Strategic Initiative:}\label{Initiative} In the previous sections we have assumed that players revise their plans until both are ready to act. Instead, we now assume that Player 1 has the ``initiative." That is, Player 1 can force the execution of plans at any iteration of the game, in an effort to catch Player 2 ``on the wrong foot." 

In addition to our baseline model's signal-plan configurations (\ref{equilibrium1})-(\ref{equilibrium4}), where both players were ready to execute their plans, there are now four additional signal-plan configurations where Player 1 forces Player 2 to act. First, there are two signal-plan configurations where Player 1 forces play, while Player 2 (correctly) thinks that he will loose. Moreover, there are two signal-plan configurations where Player 1 initiates play, with Player 2 (incorrectly) expecting defeat:
\begin{eqnarray}
&& P(\hat{A}_1,\hat{A}_2|A_1,B_2)=(1-p)q  \label{equilibrium1initiative1}\\
&& P(\hat{A}_1,\hat{B}_2|A_1,A_2)=p(1-q)  \label{equilibrium2initiative2} \\
&& P(\hat{B}_1,\hat{A}_2|B_1,B_2)=p(1-q)  \label{equilibrium3initiative3} \\
&& P(\hat{B}_1,\hat{B}_2|B_1,A_2)=(1-p)q \label{equilibrium4initiative4} \\
&& P(\hat{A}_1,\hat{B}_2|A_1,B_2)=(1-p)(1-q)\label{equilibrium4initiative5}\\
&& P(\hat{B}_1,\hat{A}_2|B_1,A_2)=(1-p)(1-q) \label{equilibrium4initiative6}\\
&& P(\hat{B}_1,\hat{B}_2|B_1,B_2)=pq \label{equilibrium4initiative7}\\
&& P(\hat{A}_1,\hat{A}_2|A_1,A_2)=pq \label{equilibrium4initiative8}\\
&& \gamma=\frac{1}{2}\big((1-p)q+p(1-q)+pq+(1-p)(1-q)\big)=\frac{1}{2} \label{equilibrium4initiative9}\\
&& \alpha_1=\alpha_2=1/2
\end{eqnarray}

\textbf{Winning Odds:} Taken together (\ref{equilibrium1initiative1})-(\ref{equilibrium4initiative9}) imply that Player 1 wins the game with probability $p>1/2$. This is intuitive given that Player 1 forces the execution of plans whenever his signal indicates a winning position. At the same time, Player 2 cannot use his information to withdraw from loosing positions. The strategic initiative thus endows Player 1 with a major advantage over Player 2.\footnote{For example, suppose $p=q=2/3$. Player 1 now wins $2/3$ of all games, while he wins only 1/2 of all games in the baseline model.} Moreover, Player 1's efforts to catch Player 2 off guard, increase $\gamma$. That is, the probability with which a battle is fought at each iteration of the game increases once one player has the ability to unilaterally initiate play.

\paragraph{Ability to Withdraw:} An edge, other than the ability to force engagements that the other player would prefer to avoid, is the ability to withdraw from battles that the opponent is willing to accept. That is, commanders of highly mobile forces, such as the Huns, who often withdrew from situations (\ref{equilibrium1initiative1})-(\ref{equilibrium4initiative4}), where the enemy was willing to fight, and forced battles in situations (\ref{equilibrium4initiative5})-(\ref{equilibrium4initiative8}) where enemy seemed ill positioned, can increase the probability of success from $p>1/2$ to $\frac{pq}{(1-p)(1-q)+pq}>p>1/2$. 


\subsection{Decoys, Deception, and Camouflage}\label{Decoy} In this section we consider games where (i) certain plans are harder to read than others, and (ii) we study the use of decoys, which players use to offset each others' scouting efforts. In particular, we find that decoys are most valuable to players facing an opponent who has what we called the ``the initiative."

\paragraph{Harder to Observe Preparations:} One may assume that the preparations associated with certain actions, e.g. action $B$, are harder to read than those associated with action $A$. In turn, players no-longer plan $A$ and $B$ with equal probability of $1/2$. Instead, they lean more towards the harder to observe actions. Analysis of this case is parallel to the settings that we already discussed, and is given in Appendix \ref{Appendix B}.

\paragraph{Decoys and Deception:} To accommodate the use of decoys, such as the fielding of rubber tanks, spoofing and jamming in air combat, or keeping the majority of one's forces hidden to lure the enemy into an ambush, we amend our model:
\begin{enumerate}
    \item \textit{Planning Stage:} Players $i=1,2$ choose the probabilities $\alpha_i$ and $1-\alpha_i$, with which they plan actions $A_i$ and $B_i$ respectively.
    \item \textit{Decoy Stage:}  Based on their planned actions, players $i=1,2$ submit decoy signals $\Check{S}_{-i}\in\{\hat{A}_{-i},\hat{B}_{-i}\}$ to their opponents. These decoy signals replace the informative signal, which players received in the baseline model, with probability $\delta$. With probability $1-\delta$, the decoy fails, and the opponent receives an informative signal which reveals the other player's true plan with probabilities $p$ and $q$, as in our baseline model. Finally, we assume that players do not know whether their adversary received the decoy or the informative signal.
    \item \textit{Execution/Revision of Plans:} Given their planned action and given their signal (which might have been a decoy), players choose whether to execute or to revise their plans. Payoffs materialize once players execute their plans.
\end{enumerate}
Graph 1 illustrates how players use decoys to degrade their adversary's information: Suppose Player 2 plans action $A_2$. In turn, with probability $1-\delta$, Player 1 receives an informative signal over his adversary's planned action just like in the baseline model. With probability $\delta$, Player 1 receives the decoy signal that Player 2 submits. Regarding this decoy, Player 2 can choose a mixed strategy. In Graph 1, Player 2 submits a decoy signal $\check{S}_1=\hat{B}_1$ to Player 1 with probability $\xi_2$. Player 2 submits $\check{S}_1=\hat{A}_1$ with probability $1-\xi_2$.\footnote{As we will see, if $\delta$ is large, players choose mixed strategies regarding the decoy signal. If $\delta$ is small, they always choose to submit misinformation, i.e. choose $\xi=1$.}
\begin{center}	
		\begin{tikzpicture}[scale=1.8]
		\tikzstyle{solid node}=[circle,draw,inner sep=2,fill=black]
		\tikzstyle{solid node sq}=[rectangle,draw,inner sep=2,fill=black]
		\tikzstyle{hollow node}=[circle,draw,inner sep=2]
		\tikzstyle{empty node}=[draw=none,fill=none]
		\tikzstyle{level 1}=[level distance=15mm,sibling distance=5cm]
		\tikzstyle{level 2}=[level distance=15mm,sibling distance=2.5cm]
		\tikzstyle{level 3}=[level distance=15mm,sibling distance=1.5cm]
		\tikzstyle{level 4}=[level distance=15mm,sibling distance=2cm]
		\tikzstyle{level 5}=[level distance=15mm,sibling distance=1.8cm]
		\tikzstyle{level 6}=[level distance=15mm,sibling distance=0.9cm]
		\node(0)[hollow node,label=above:{$A_2$}]{}
		child{node[hollow node,label=above left:{}]{}
			child{node[empty node,label=below:{$\hat{A}_1$}]{} edge from parent node[left]{$p$}}
			child{node[empty node,label=below right:{$\hat{B}_1$}]{}
				edge from parent node[right]{$1-p$}}
			edge from parent node[left,xshift=-5]{$1-\delta$}
		}
		child{node[hollow node,label=above right:{}]{}
			child{node[empty node,label=below left:{$\hat{A}_1$}]{}
				edge from parent node[left]{$1-\xi_2$}}
			child{node[empty node,label=below:{$\hat{B}_1$}]{} edge from parent node[right]{$\xi_2$}}
			edge from parent node[right,xshift=5]{$\delta$}};
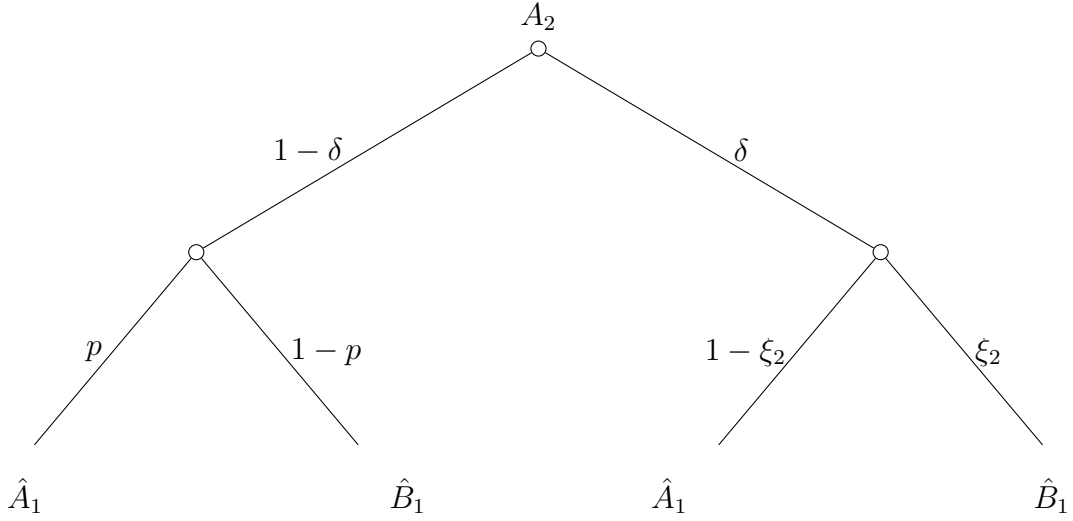
		\end{tikzpicture}\\Graph 1: Information with Decoys. Player 1's signals conditional on Player 2's plan to play action $A_2$.
		
	\end{center}
 
\paragraph{Optimal use of Decoys:} Following Graph 1, it is useful to define $\tilde{p}:=P(\hat{A}_1|A_2)=P(\hat{B}_1|B_2)=\delta(1-\xi_2)+(1-\delta)p$ and $\tilde{q}:=P(\hat{A}_2|A_1)=P(\hat{B}_2|B_1)=\delta(1-\xi_1)+(1-\delta)q$, such that $\tilde{p}$ and $\tilde{q}$ are the probabilities with which players read each others' plans and preparations correctly. 
In turn, players 1 and 2 can choose the probabilities $\xi_1,\xi_2$, to minimize the probabilities $\tilde{p},\tilde{q}$ with which their adversary can read their plans correctly. That is, Player 2 chooses $\xi_2=\frac{P-1/2}{\delta}+1-p$ as long as $\frac{P-1/2}{\delta}+1-p\in[0,1]$. Otherwise, if $\frac{P-1/2}{\delta}+1-p>1$, we have $\xi_2=1$, and if $\frac{P-1/2}{\delta}+1-p<0$, we have $\xi_2=0$.\footnote{Put differently, to minimize the precision with which Player 1 can infer Player 2's plans, Player 2 chooses $\xi$ such that $P(\hat{A}_1|A_2)=\delta(1-\xi)+(1-\delta)p=P(\hat{B}_1|A_2)=\delta\xi+(1-\delta)(1-p)=1/2$. That is, if $\delta$ is large, Player 2 has an incentive to submit true, rather than false information, to Player 1 some of the time. In the extreme case $\delta=1$, the optimal strategy, which minimizes the other player's information is thus to send true (and false) information with probability $1/2$.}

Substituting the new signal ``precisions" $\tilde{p}$ and $\tilde{q}$, we can once again use equations (\ref{equilibrium1})-(\ref{alpha}) and (\ref{equilibrium1initiative1})-(\ref{equilibrium4initiative9}), allows to compute how the use of decoys impacts winning odds in the models discussed above. Doing so reveals that decoys are most valuable for players who face an adversary who has what we called ``the initiative." That is, if Player 1 has the initiative, he will win (lose) whenever he correctly (incorrectly) reads his adversary's plan. In turn, Player 2 can use his decoy to directly degrade the winning probability from $p$ to $\tilde{p}=\delta(1-\xi_2)+(1-\delta)p$, where $\xi_2=\frac{P-1/2}{\delta}+1-p$.

Finally, we note that the above calculations on the use of decoys assumed that Player 1 rationally anticipates Player 2's use of decoys, such that the best that Player 2 can do is to minimize Player 2's wining probability to $1/2$. If Player 1 didn't know that Player 2 was using a decoy, the winning probability for Player 1 would be less than $1/2$. That is, an unsuspecting player, who tries to exploit his ability to force engagements, is most vulnerable to the use of decoys.

\section{Discussion}\label{Discussion}

The present paper stresses that (i) every action requires preparation and (ii) that these preparations are being monitored such that (iii) players partially know each others' contemporaneous actions once they decide to execute their plans. The current approach thus avoids the ``chicken egg" type of problems, between signal and action, that emerge in standard simultaneous move games where players observe each other.\footnote{See \citet{Ang06}, p. 1730, and \citet{Rub89} for a discussion of the difficulties that prevent the study of one-shot simultaneous move games where players observe each others' contemporaneous actions.}

   
Once players monitor each others' preparations, equilibrium play differs from standard matching pennies in several ways. First, rather than relying on a 50-50 coin toss, players screen for favorable engagements, and they only act once their information over their adversary's planned action indicates a  ``winning position." In turn, rather than the strongest or fastest player, it is the player who best reads his adversary that wins the game.\footnote{Unlike the standard matching pennies game, the present approach thus accommodates the views of practitioners such as Andre Agassi and Boris Becker, which we alluded to in the introduction. Likewise Mario Gomez recalls a conversation with Peter Czech:``...He basically told me what I was thinking [while preparing the penalty kick]... that's when I understood what a great goal keeper he is... reading the players and the players' body-language."} Second, the probability that players execute their plans at each iteration of the game is decreasing in the precision with which they monitor each other. Put differently, whenever both players read their adversary's plans well,they will revise their plans more often, and are thus more reluctant to engage in conflict at each iteration of the game. Finally, players, who do not fully understand equilibrium, overestimate the probability with which they will win an engagement. 

We extended our model in several directions, and studied (i) the use of decoys and misinformation as well as (ii) environments where one player has the ``strategic initiative," which allows him to force an engagement that the other player would like to avoid. These extensions indicate that conflict unfolds more rapidly once one player has the strategic initiative, and that the use of decoys and misinformation were the best tools to counter and adversary who has the strategic initiative.

\newpage

\addcontentsline{toc}{section}{References}
\markboth{References}{References}
\bibliographystyle{apalike}
\bibliography{References}

\newpage
\begin{section}{Appendix A}\label{Appendix A}
    \begin{eqnarray}
   && P(A_1,A_2)=P(\hat{A}_1,\hat{B}_2|A_1,A_2)\alpha_1\alpha_2= p(1-q)\alpha_1\alpha_2
    \label{equilibrium1q}\\
   && P(B_1,B_2)=P(\hat{B}_1,\hat{A}_2|B_1,B_2)(1-\alpha_1)(1-\alpha_2)= p(1-q)(1-\alpha_1)(1-\alpha_2)  \label{equilibrium2q} \\
   && P(A_1,B_2)=P(\hat{A}_1,\hat{A}_2|A_1,B_2)\alpha_1 (1-\alpha_2)= (1-p)q\alpha_1 (1-\alpha_2) \label{equilibrium3q} \\
   && P(B_1,A_2)=P(\hat{B}_1,\hat{B}_2|B_1,A_2)(1-\alpha_1)\alpha_2=  (1-p)q(1-\alpha_1)\alpha_2     \label{equilibrium4q}
\end{eqnarray}
Where $\alpha_1$ and $\alpha_2$ are the probabilities with which players $1$ and $2$ play actions $A_1$ and $A_2$ respectively.
\begin{eqnarray}
    \max_{\alpha_1}E_1[U_1]=&&(-(1-p)q\alpha_1 (1-\alpha_2)+p(1-q)\alpha_1\alpha_2\\&&+p(1-q)(1-\alpha_1)(1-\alpha_2)-(1-p)q(1-\alpha_1)\alpha_2)\nonumber
\end{eqnarray}
\begin{eqnarray}
    \max_{\alpha_2}E_2[U_2]=&&(+(1-p)q \alpha_1 (1-\alpha_2)-p(1-q) \alpha_1\alpha_2\\&&-p(1-q) (1-\alpha_1)(1-\alpha_2)+(1-p)q (1-\alpha_1)\alpha_2)\nonumber
\end{eqnarray}
Solving FOCs for $\alpha_1,\alpha_2$ yields:
\begin{align}
\alpha_{1}&=\frac{(1-p)q+p(1-q)}{q(1-p)+q(1-p)+p(1-q)+p(1-q)}=1/2\label{alpha1}\\
\alpha_{2}&=\frac{(1-p)q+p(1-q)}{q(1-p)+q(1-p)+p(1-q)+p(1-q)}=1/2\label{alpha2}
\end{align}
\end{section}

\begin{section}{Appendix B}\label{Appendix B}
Suppose that e.g. player 2's action $B_2$ is harder to observe than action $A_2$. That is, given that player 2 plays $B_2$, player 1 observes this action with probability $\tilde{p}$, where $\tilde{p}<p$.
\begin{eqnarray}
   && P(A_1,B_2)=P(\hat{A}_1,\hat{A}_2|A_1,B_2)\alpha_1 (1-\alpha_2)= (1-\tilde{p})q\alpha_1 (1-\alpha_2)
    \label{equilibrium1q}\\
   && P(A_1,A_2)=P(\hat{A}_1,\hat{B}_2|A_1,A_2)\alpha_1\alpha_2= p(1-q) \alpha_1\alpha_2 \label{equilibrium2q} \\
   && P(B_1,B_2)=P(\hat{B}_1,\hat{A}_2|B_1,B_2)(1-\alpha_1)(1-\alpha_2)= \tilde{p}(1-q)(1-\alpha_1)(1-\alpha_2) \label{equilibrium3q} \\
   && P(B_1,A_2)=P(\hat{B}_1,\hat{B}_2|B_1,A_2)(1-\alpha_1)\alpha_2=  (1-p)q(1-\alpha_1)\alpha_2     \label{equilibrium4q}
\end{eqnarray}
Where $\alpha_1$ and $\alpha_2$ are the probabilities with which players $1$ and $2$ play actions $A_1$ and $A_2$ respectively.
\begin{eqnarray}\max_{\alpha_1}E_1[U_1]=&&(-(1-\tilde{p})q\alpha_1 (1-\alpha_2)+p(1-q)\alpha_1\alpha_2\\&&+\tilde{p}(1-q)(1-\alpha_1)(1-\alpha_2)-(1-p)q(1-\alpha_1)\alpha_2)\nonumber\end{eqnarray}
\begin{eqnarray}\max_{\alpha_2}E_2[U_2]=&&(+(1-\tilde{p})q\alpha_1 (1-\alpha_2)-p(1-q)\alpha_1\alpha_2\\&&-\tilde{p}(1-q)(1-\alpha_1)(1-\alpha_2)+(1-p)q(1-\alpha_1)\alpha_2)\nonumber\end{eqnarray}
Solving FOCs for $\alpha_1,\alpha_2$ yields:
\begin{align}\alpha_{1}&=\frac{\tilde{p}(1-q)+(1-p)q}{(1-p)q+p(1-q)+\tilde{p}(1-q)+(1-p)q}\label{alpha1}\\
\alpha_{2}&=\frac{\tilde{p}(1-q)+(1-\tilde{p})q}{(1-\tilde{p})q+p(1-q)+\tilde{p}(1-q)+(1-p)q}\label{alpha2}
\end{align}
Comparison of (\ref{alpha1}) and (\ref{alpha2}), recalling $\tilde{p}<p$ indicates that $\alpha_2<\alpha_1$.
\end{section}

\end{document}